\documentclass[12pt,preprint]{aastex}

\newcommand{\sfrd}{$\rho_{\star}$}
\newcommand{\sfrduv}{$\rho_{\star\rm (UV)}$}
\newcommand{\sfrdrad}{$\rho_{\star\rm (rad)}$}

\slugcomment{To appear in AEGIS Special Issue of ApJ Letters}

\shorttitle{AEGIS20: a radio survey of the Extended Groth Strip}
\shortauthors{Ivison et al.}

\begin{document}

\title{AEGIS20: a radio survey of the Extended Groth Strip}

\author{R.\,J.\ Ivison,\altaffilmark{1}
S.\,C.\ Chapman,\altaffilmark{2} S.\,M.\ Faber,\altaffilmark{3}
Ian Smail,\altaffilmark{4} A.\,D.\ Biggs,\altaffilmark{1}
C.\,J.\ Conselice,\altaffilmark{5}\\ G.\ Wilson,\altaffilmark{6}
S.\ Salim,\altaffilmark{7} J.-S.\ Huang\altaffilmark{8}
\& S.\,P.\ Willner\altaffilmark{8}
}

\altaffiltext{1}{UK Astronomy Technology Centre, Royal Observatory,
                 Blackford Hill, Edinburgh EH9\,3HJ, UK}
\altaffiltext{2}{Astronomy Department, California Institute of
                 Technology, Pasadena, CA\,91125}
\altaffiltext{3}{Department of Astronomy and
                 Astrophysics, University of California, Santa Cruz,
                 Santa Cruz, CA\,95064}
\altaffiltext{4}{Institute for Computational Cosmology, Durham
                 University, South Road, Durham DH1\,3LE, UK}
\altaffiltext{5}{Physics and Astronomy, University of
                 Nottingham, University Park, Nottingham NG7\,2RD, UK}
\altaffiltext{6}{Spitzer Science Center, California Institute of Technology,
                 1200 E.\ California Bvd, Pasadena, CA\,91125}
\altaffiltext{7}{Department of Physics and Astronomy, University of California,
                 Los Angeles, CA\,90095}
\altaffiltext{8}{Harvard-Smithsonian Center for Astrophysics, 60 Garden Street,
                 Cambridge, MA\,02138}

\setcounter{footnote}{8}

\begin{abstract}
We describe AEGIS20 -- a radio survey of the Extended Groth Strip
(EGS) conducted with the Very Large Array (VLA) at 1.4\,GHz. The
resulting catalog contains 1,123 emitters and is sensitive to
ultraluminous (10$^{12}$\,L$_{\odot}$) starbursts to $z\le\rm 1.3$,
well matched to the redshift range of the DEEP2 spectroscopic survey
in this region. We use stacking techniques to explore the
$\mu$Jy-level emission from a variety of galaxy populations selected
via conventional criteria -- Lyman-break galaxies (LBGs), distant red
galaxies (DRGs), UV-selected galaxies and extremely red objects (EROs)
-- determining their properties as a function of color, magnitude and
redshift and their extinction-free contributions to the history of
star formation. We confirm the familiar pattern that the
star-formation-rate (SFR) density, \sfrd, rises by {\em at least}
$\sim$5$\times$ from $z$ = 0 to 1, though we note highly discrepant
UV- and radio-based SFR estimates. Our radio-based SFRs become more
difficult to interpret at $z>\rm 1$ where correcting for contamination
by radio-loud active galactic nuclei (AGN) comes at the price of
rejecting luminous starbursts. Whilst stacking radio images is a
useful technique, accurate radio-based SFRs for $z\gg\rm 1$ galaxies
require precise redshifts and extraordinarily high-fidelity radio data
to identify and remove accretion-related emission.
\end{abstract}


\section{Introduction}

The tight correlation between radio and far-IR emission for
star-forming galaxies (Helou et al.\ 1985; Garrett 2002; Kovacs et
al.\ 2006), allows us to push dust-independent surveys down to lower
SFRs than is possible in the confusion-limited far-IR/submm
wavebands. Moreover, the high mapping speed of facilities such as the
Giant Metre-wave Radio Telescope means we can quickly obtain the large
samples of faint sources needed for reliable analyses.

The bulk of the far-IR background seen by {\it COBE} (Fixsen et al.\
1998) most likely arises from a large population of luminous and
ultraluminous IR galaxies (LIRGs and ULIRGs), their energy originating
from dust-obscured star formation and accretion. Individually less
luminous than submm galaxies, with $L_{\rm bol}\sim 3\times
10^{11}$\,L$_{\odot}$, these galaxies are believed to be sufficiently
numerous to dominate \sfrd\ at $z\sim$ 1 (Dole et al.\ 2006).

In this letter we present a new panoramic radio survey -- AEGIS20 --
undertaken with the National Radio Astronomy
Observatory's\footnote{The National Radio Astronomy Observatory is
operated by Associated Universities Inc., under a cooperative
agreement with the National Science Foundation.} VLA as part of the
All-wavelength Extended Groth Strip International Survey (AEGIS; Davis
et al.\ 2006). AEGIS20 was tuned to detect ULIRGs robustly at $z\sim$
1, with a noise level of 10\,$\mu$Jy\,beam$^{-1}$ at 1.4\,GHz (cf.\
Hopkins et al.\ 2003; Bondi et al.\ 2003). The resulting catalog,
available electronically, contains $\sim$10$^3$ faint radio sources --
an order of magnitude more than the 5-GHz survey of this region by
Fomalont et al.\ (1991); nearly half are expected to have optical
spectra provided by the DEEP2 survey, many with redshifts, as well as
photometry across a wide range of wavelengths.

The future goal of AEGIS20 is to measure the 1.4-GHz luminosity
function, track the evolution of SFRs in LIRGs and ULIRGs and, using a
measure of the local galaxy density of each radio source, study the
history of star formation as a function of environment. Here, we
present the AEGIS20 catalog and utilise the radio image to estimate
SFRs for a number of independent and overlapping galaxy populations
selected via conventional criteria.

\section{Observations and data reduction}

Data were obtained at 1.4\,GHz during 2003--05 with the VLA in its B
configuration, acquiring seven 3.125-MHz channels every 5\,s in each
of four IFs. We obtained data in six positions, spaced by 15$'$ (see
Davis et al.\ 2006), concentrating in the northern half of the EGS
because of the proximity of 3C\,295 ($S_{\rm 1.4GHz}$ =
23\,Jy). Around 18\,hr of data were acquired for each of the field
positions, cycling through them between scans of 1400+621 and 1419+543
to monitor bandpass, amplitude and phase. Absolute flux calibration
was set using 3C\,286.

Calibrated visibilities and associated weights were used to generate
mosaics of $\rm 37\times 512^2\times 0.8''^2$-pixel images to quilt
the VLA's primary beam in each EGS field position. {\sc clean} boxes
were placed tightly around all sources and a series of {\sc imagr} and
{\sc calib} tasks were run, clipping the $uv$ data after subtracting
{\sc clean} components generated by the third iteration of {\sc
imagr}. The central images from each of the pointings were then
knitted together using {\sc flatn}, ignoring data beyond the primary
beam's half-power point, to produce a large mosaic. The synthesized
beam is circular, with a {\sc fwhm} of $\sim$3.8$''$.

\section{Sample definition}

To define a sample of radio sources we searched signal-to-noise (S/N)
images using the {\sc sad} detection algorithm, emulating the
technique described by Biggs \& Ivison (2006). Sources with
$\ge$4$\sigma$ peaks were fitted with 2-D Gaussians using {\sc jmfit},
those with $\ge$5$\sigma$ peaks surviving to be fitted in total
intensity. Sources with sizes equal to or smaller than the restoring
beam were considered unresolved; their size was constrained to that of
the beam. We make no correction for bandwidth smearing in the catalog:
this is a small effect ($\sim$5\%) given our mosaicing strategy and
the use of B configuration. We detect 38, 79, 171, 496 and 1,123
sources with $S_{\rm 1.4GHz}\ge\rm 2,000$, $\ge$800, $\ge$320,
$\ge$130 and $\ge$50\,$\mu$Jy (consistent with Simpson et al.\ 2006),
where the 5$\sigma$ detection limits at 130 and 50\,$\mu$Jy cover 0.73
and 0.04\,deg$^2$. Confusion is not an issue: the source density on an
arcmin$^2$ scale is $<$0.01\,beam$^{-1}$.

$S_{\rm 1.4GHz}=\rm 50\,\mu Jy$ corresponds to rest-frame 1.4-GHz
luminosities, $L_{\rm 1.4GHz}$, of 0.44, 2.3, 6.0 and 12 $\times
10^{23}$\,W\,Hz$^{-1}$ and SFRs of 50, 275, 725 and
1,430\,M$_{\odot}$\,yr$^{-1}$ at $z$ = 0.5, 1.0, 1.5 and 2.0 (for
$\Omega_m=0.27, \Omega_\Lambda=0.73$,
$H_0=71$\,km\,s$^{-1}$\,Mpc$^{-1}$ --- Spergel et al.\ 2003 --- a
Salpeter initial mass function [IMF] with d$N$/d$M\propto M^{-2.35}$
over 0.1--100\,M$_{\odot}$ and $S_{\nu}\propto \nu^{-0.8}$).

AEGIS20 covers 57\% of the existing DEEP2 region (Davis et al.\ 2006),
with $\sim$7,900 unique redshifts available in the 0.28\,deg$^2$
common to both surveys. DEEP2 thus covers 35\% of AEGIS20, although
its $BRI$ imaging covers 90\% of AEGIS20 (93\% of cataloged
sources). Of the AEGIS20 sources with optical imaging, $\sim$36\% have
$R_{\rm AB}<\rm 24.1$ counterparts within 1$''$. Since the DEEP2
targeting rate is $\sim$70\%, the inclusion rate on DEEP2 masks for
faint radio emitters is $\sim$25\%. At present, $\sim$100 of the
targeted AEGIS20 radio sources have DEEP2 redshifts --- a very high
success rate. Radio properties of DEEP2 galaxies and the spectroscopic
properties of the AEGIS20 catalog will be discussed in a forthcoming
paper.

\section{The radio properties of distant galaxy populations}

The wealth of multi-frequency data in AEGIS allows us to mimic the
selection of galaxy populations such as DRGs ($J-K>\rm 2.3$; expected
to lie at $\rm 1.9 < z < 3.5$ --- Franx et al.\ 2003), as well as LBGs
(Steidel et al.\ 2003) and EROs. We investigate the radio properties
of several such populations in this section, taking them roughly in
order of increasing redshift.

We expect to detect only a small fraction of distant galaxies at radio
frequencies. In such situations it is common to assess the emission
from a galaxy population using a stacking analysis, accomplished
either by extracting and co-adding postage stamps centered on the
galaxies of interest (`image stacking') or by co-adding flux densities
measured at the positions of the galaxies (`pixel stacking'). We adopt
both approaches here.  To determine the signal lost by pixel stacking
we employed radio emitters with S/N = 5--20 pixel$^{-1}$, finding a
difference of only 3.9\% between the values returned at the positions
of the emitters and cataloged AEGIS20 flux densities. Monte-Carlo
simulations show that the mean $S_{\rm 1.4GHz}$ determined by pixel
stacking are slightly skewed ($+$0.1\,$\mu$Jy, typically) but are
otherwise well described by Gaussian statistics; medians are affected
at the $<$0.01-$\mu$Jy level. $S_{\rm 1.4GHz}$ values have been
corrected for bandwidth smearing (+5.0\%), for pixel-stacking losses
(+3.9\%) and we have excluded galaxies in noisy regions ($\sigma_{\rm
1.4GHz}>\rm 30\,\mu Jy\,beam^{-1}$).

We must excise emission due to accretion if we are to determine
accurate radio-based SFRs. Morphological classification of most radio
emitters is not feasible at the resolution of our data, spectral
indices are not to hand and the availability and reliability of AGN
indicators at shorter wavelengths differs widely across the
EGS. Radio-loud AGN were thus identified and rejected via a $L_{\rm
1.4GHz}$ limit. Following Condon (1992), we adopt $L_{\rm 1.4GHz} <\rm
10^{24}\,W\,Hz^{-1}$ for normal galaxies, an order of magnitude below
the break in morphology and luminosity noted by Fanaroff \& Riley
(1974). We quote the noise-weighted mean $S_{\rm 1.4GHz}$; where AGN
contamination is extreme ($>$5\%), we quote the median, noting the
number of obvious AGN. One unfortunate consequence of excising
radio-loud AGN on the basis of $L_{\rm 1.4GHz}$ is the exclusion of
distant hyperluminous starbursts lying on the far-IR/radio correlation
(\S5).

We begin with a ultraviolet (UV)-selected catalog containing 4,426
galaxies detected at 230\,nm by {\em GALEX}, with DEEP2 redshifts,
i.e.\ $R_{\rm AB}<\rm 24.1$, excluding objects with AGN flags (Salim
et al.\ 2006). Of these, 3,908 lie within the 0.28\,deg$^2$ of common
areal coverage with AEGIS20.  We compare SFRs determined in two ways
-- via their UV and radio properties, SFR$_{\rm UV}$ and SFR$_{\rm
rad}$ -- for the same galaxies. We use UV-based, extinction-corrected
SFRs, derived by comparing observed spectral energy distributions
(SEDs) with those of model galaxies exhibiting a wide range of
properties and SF histories (Salim et al.\ 2005). Appropriate volume
corrections for the {\em GALEX}-selected sample are difficult to
determine due to a complex selection function dependent on UV/optical
magnitudes and spectral characteristics. We therefore correct for the
rate with which {\em GALEX} detects DEEP2 galaxies, which is known to
fall from 90--75--60\% at $z$ = 0.2--1.0--1.4, but not for our
steadily decreasing sensitivity to low-luminosity galaxies (Arnouts et
al.\ 2005). Because of this, we are limited to discussing the ratio of
\sfrduv\ and \sfrdrad\ within the sample.

Robust radio detections of the UV sample were possible by pixel
stacking over $\Delta z=\rm 0.2$ bins. Only 11 radio-loud AGN were
identified via $L_{\rm 1.4GHz}$, $<<$1\% of the total; having rejected
these, noise-weighted means provide the most appropriate measure of
SFR for this sample (Table~\ref{table}). SFR$_{\rm UV}$ and SFR$_{\rm
rad}$ per UV-selected galaxy both increase with redshift,
unsurprisingly since we are probing more UV-luminous galaxies at
larger distances. For $z$ = 0--1, \sfrduv\ remains fairly constant
whilst \sfrdrad\ rises rapidly. It may seem puzzling that \sfrduv\ at
$z\sim\rm 0$ is an order of magnitude higher than \sfrdrad\
(Fig.~\ref{sfrd}) --- \sfrdrad\ should be sensitive to all recent star
formation, obscured and unobscured, for a constant IMF --- however,
Bell (2003) showed that $L_{\rm far-IR}/L_{\rm UV}$ varies by
$\ga$30$\times$ between 0.01 and 3\,L$_{\star}$ and that radio data
underestimate SFRs in low-luminosity galaxies typical of those
detected locally by {\em GALEX}. Hopkins \& Beacom (2006) argue that
for the full picture we should add SFR$_{\rm UV}$ and SFR$_{\rm rad}$.
\sfrdrad\ and \sfrduv\ achieve parity at $z\sim\rm 0.4$, after which
\sfrdrad\ continues to rise until $z\sim\rm 0.7$ (cf.\ Cowie et al.\
2004) when incompleteness seriously impacts the sample. Although it
is tempting to speculate that the rise in \sfrdrad\ results from the
increasing dominance of dust-obscured IR-luminous galaxies, we must
recall our sample's origins. We are witnessing an increasing SFR per
{\em UV-selected} galaxy, partly because at $z\sim\rm 1$ we are
probing the most UV-luminous galaxies; we are also witnessing an
increase in \sfrdrad\ {\em despite} the increasing
incompleteness. Adding SFR$_{\rm UV}$ and SFR$_{\rm rad}$, \sfrd\ due
to UV-selected galaxies increases as {\em at least} (1+$z$)$^{2.2}$
between $z$ = 0 and 1 (cf.\ Schiminovich et al.\ 2005).

The mismatch between the absolute and relative rates of SF derived
using UV- and radio-based indicators is worrying, particularly the
difference between local estimates of \sfrd.  The local \sfrduv\
matches the \sfrd\ compilation presented by Hopkins \& Beacom (2006),
which implies the UV-selected sample accounts for most of the SF in
the local Universe, yet the UV sample at 0.0\,$<$\,$z$\,$<$\,0.2
accounts for $\ll$1\% of the total cataloged $S_{\rm 1.4GHz}$ in the
region of common areal coverage. If the fraction of $S_{\rm 1.4GHz}$
due to obscured SF at 0.0\,$<$\,$z$\,$<$\,0.2 exceeds 1\%, \sfrd\
would then be at the upper envelope of commonly accepted values.
 
Moving to slightly higher redshifts, we take two catalogs of EROs.
One uses the conventional color cut $R-K_{\rm s}>\rm 5.3$ with $K_{\rm
s}<\rm 20.5$ and DEEP2 redshifts, 1.0\,$<$\,$z$\,$<$\,1.5 (Conselice et
al., in prep). It contains 382 objects in low-noise areas of our radio
image. The second (Wilson et al.\ 2006) uses $R_{\rm AB}-3.6\mu\rm m$
$>$ 4 to select essentially the same class of objects, but the larger
area covered at 3.6\,$\mu$m (and no requirement for redshifts) yields
2,363 objects in 0.26\,deg$^2$ of AEGIS20. We assumed $z=\rm 1.1$ to
excise radio-loud AGN from this sample.

Both ERO samples are well detected at 1.4~GHz, as shown in Fig.~2. In
the spectroscopic sample, mean $S_{\rm 1.4GHz}$ does not vary
significantly as a function of color, though it is a function of
$K_{\rm s}$ consistent with the findings of Smail et al.\ (2002).
Median $L_{\rm 1.4GHz}$ is $\rm 8.1\times 10^{22}$\,W\,Hz$^{-1}$ and
the median SFR per ERO is 92\,$\pm$\,7\,M$_{\odot}$\,yr$^{-1}$.  In
the sample volume set by the redshift limits and survey area this
equates to \sfrd\ = 0.07\,M$_{\odot}$\,yr$^{-1}$\,Mpc$^{-3}$ (cf.\
Georgakakis et al.\ 2006; Simpson et al., in prep).

The larger ERO sample reveals a weak trend for $S_{\rm 1.4GHz}$ to
increase with redness; this is confirmed by the increasing detection
rate for individual objects (Table~1).  Mean $S_{\rm 1.4GHz}$ also
declines as $S_{\rm 3.6 \mu m}$ decreases.  We expect $S_{\rm 3.6 \mu
m}$ to trace stellar mass and distance, and the factor of $\sim$5.5
decrease in $S_{\rm 1.4GHz}$ for a for a factor of $\sim$15 decrease
in $S_{\rm 3.6 \mu m}$ suggests an increasing SFR per unit stellar
mass as redshift increases. The overall \sfrd\ for this sample is
consistent with that of the spectroscopic sample, as expected given
the limited number of spectroscopic redshifts and the significant
sample overlap.

Huang et al.\ (in prep) present a catalog selected at $S_{\rm 24\mu
m}>\rm 150\mu$Jy, with $S_{\rm 4.5\mu m}>S_{\rm 3.6\mu m}$, aiming to
select galaxies and AGN at $z>$ 1.5. Almost 10$^3$ objects lie in
low-noise regions of our radio mosaic, overlapping AEGIS20 by
0.26\,deg$^2$. The individual radio detection rate is a strong
function of $S_{\rm 24\mu m}$, rising from 30 to 70\% between $S_{\rm
24\mu m}=$ 0.15 and $>$1.2\,mJy. Median $S_{\rm 1.4GHz}$
(Table~\ref{table}) is fairly insensitive to $R_{\rm AB}$, varying by
$<$2$\times$ over $>$3 mag. Over 40\% of the 24-$\mu$m-selected
galaxies have $L_{\rm 1.4GHz}>\rm 10^{24}$\,W\,Hz$^{-1}$ when assuming
$z=\rm 2.75$. At this redshift {\em all} S/N $\ge$ 3 measurements
imply radio-loud AGN and it is difficult to estimate the SFR: \sfrd\
is likely to be high, but so is the level of accretion-related
contamination. The median $S_{\rm 1.4GHz}$, 28\,$\mu$Jy, translates
into \sfrd\ = 0.20\,M$_{\odot}$\,yr$^{-1}$\,Mpc$^{-3}$ for $z$ =
1.5--4. We would be unsurprised if this is in error by $\times$2;
regardless, this is an important star-forming population.

Moving on to yet more distant populations, this time to a sample of
DRGs selected at $K_{\rm s}<\rm 20.5$ (Vega) with $J-K_{\rm s}>\rm
2.3$ (Conselice et al.\ 2006), 108 of which lie within the
0.11\,deg$^2$ of common coverage with AEGIS20.  Although expected to
lie at 1.9\,$<$\,$z$\,$<$\,3.5 (Franx et al.\ 2003), Conselice et al.\
find that 64\% lie at 1\,$<$\,$z$\,$<$\,2. Galaxies with $z<\rm 1$,
evident via DEEP2, have been removed from the sample used here. One of
the radio emitting DRGs has $L_{\rm 1.4GHz}$ consistent with
radio-loud AGN (or, as noted earlier, a hyperluminous starburst). The
mean $S_{\rm 1.4GHz}$ for the DRGs was $\rm 10.1\pm 1.3\,\mu$Jy --
faint emission can be seen in the stacked S/N image
(Fig.~\ref{stack}). At $z=\rm 1.5$ this corresponds to $L_{\rm 1.4GHz}
= (1.2\pm 0.2)\times 10^{23}$\,W\,Hz$^{-1}$ and a mean SFR (per DRG)
of $\rm 150\pm 19$\,M$_{\odot}$\,yr$^{-1}$. Knudsen et al.\ (2005)
found $\rm 190\pm 50$\,M$_{\odot}$\,yr$^{-1}$ using submm data for a
sample of 30 DRGs (adapting to the cosmology and IMF used here),
having assumed significantly larger distances. The observed radio
emission from $K_{\rm s}<\rm 20.5$ DRGs equates to \sfrd\ =
0.02\,M$_{\odot}$\,yr$^{-1}$\,Mpc$^{-3}$ at $z$ = 1--2.

The LBGs of Steidel et al.\ (2003) lie in a noisy region of the radio
mosaic. Of the 334 cataloged LBGs, after correction for the
astrometric offset in that catalog ($\Delta\alpha=\rm +0.8'',
\Delta\delta=\rm +2.6''$), 107 lie within low-noise regions of the
radio mosaic; their mean $S_{\rm 1.4GHz}$ was $\rm 2.0\pm 2.3\,\mu$Jy
(median, $-$0.6\,$\mu$Jy), consistent with an average SFR of
$<$500\,M$_{\odot}$\,yr$^{-1}$ (3\,$\sigma$, for $z=\rm
3$). Restricting the catalog to the 53 LBGs detected at 8\,$\mu$m with
IRAC did not change the situation significantly (cf.\ Rigopoulou et
al.\ 2006).

Finally, Huang et al.\ (2005) describe a population of IR-luminous
LBGs (ILLBGs) detected at $S_{\rm 24\mu m}>\rm 60\mu$Jy. Only six of
Huang et al.'s 13 objects lie within our radio mosaic. Their median
$S_{\rm 1.4GHz}$ is 44.2\,$\mu$Jy, including the one significant
detection: Westphal MD99 at $\sim$1\,mJy. This provides tentative
support for the assertion that ILLBGs share the high SFRs of submm
galaxies, though this is a very small sample in a particularly noisy
region of the radio mosaic and accretion-related contamination is
possible. If, as Huang et al.\ suggest, ILLBGs lie at
2\,$<$\,$z$\,$<$\,3 (like submm galaxies -- Chapman et al.\ 2005) then
their \sfrd\ is similar to that of the 24$\mu$m-selected galaxies with
which they will overlap significantly (Fig.~\ref{sfrd}).

\section{On radio data as a probe of global SF history}

\begin{figure}
\epsscale{0.5}
\plotone{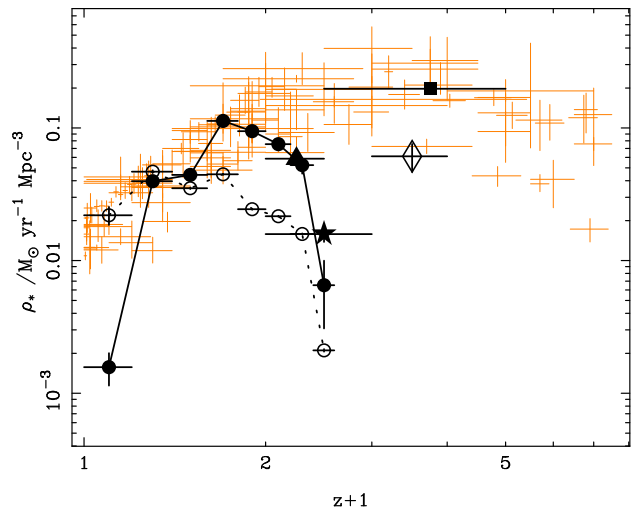}
\caption{Radio-based estimates of SFR density (\sfrd) for a number of
independent and overlapping galaxy populations selected via magnitude
and color criteria: UV-selected galaxies (open and filled circles for
UV- and radio-based SFRs, joined by dotted and solid lines), $R-K_{\rm
s}>\rm 5.3$ EROs (triangle), DRGs (star), ILLBGs (diamond) and
24-$\mu$m-selected galaxies (filled square). These are conservative
estimates -- no attempt has been made to correct for accessible
volume; contamination by radio-loud AGN is possible at $z\gg\rm 1$
(\S5).  The upper envelope of points thus traces the minimum \sfrd\ as
a function of redshift, as demonstrated by the compilation of \sfrd\
data from Hopkins \& Beacom (2006), plotted faintly here.
\label{sfrd}}
\end{figure}

\begin{figure}
\epsscale{0.5}
\plotone{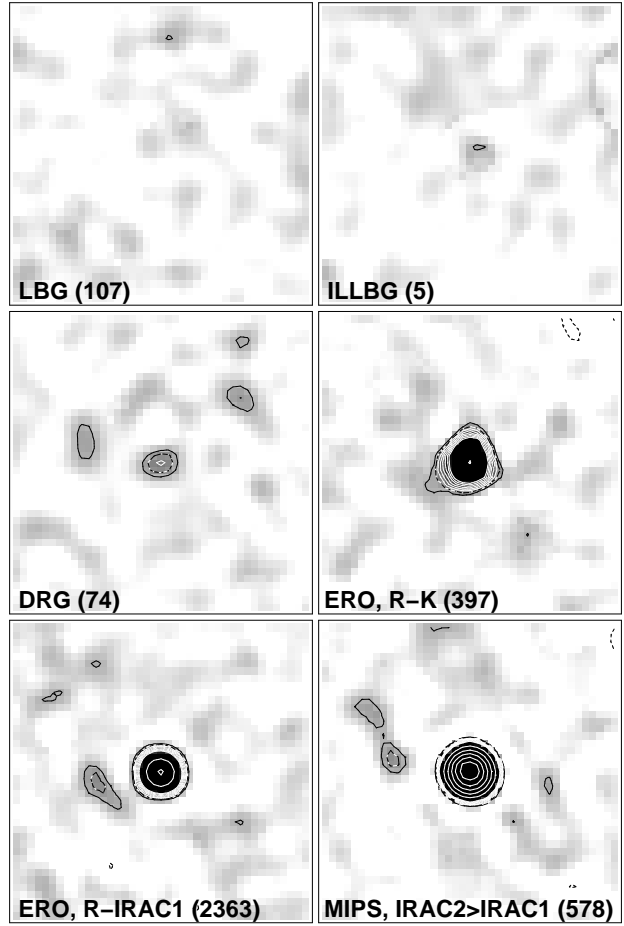}
\caption{Stacked S/N images ($\rm 33''\times 33''$) at the positions
of the galaxies described in \S4: conventional and IR-luminous LBGs
(top row); DRGs and $R-K_{\rm s}>\rm 5.3$ EROs (middle row);
IRAC-selected EROs and MIPS-selected galaxies with $S_{\rm 4.5\mu
m}>S_{\rm 3.6\mu m}$ (bottom row). Areas of high noise and galaxies
detected individually ($\ge$5$\sigma$) were excluded. Contours are
plotted at S/N levels of $-3,3,4...10,20...100$ and the greyscale is
identical in each case. The number of stacked sources from each
population is shown in parentheses. The detection of the IRAC-selected
ERO population is so significant that we see
secondary structure: the image resembles the dirty beam
since none of the individually undetected sources in the ensemble that
makes up the stacked image have been {\sc clean}ed.
\label{stack}}
\end{figure}

Fig.~\ref{sfrd} shows \sfrd\ for the galaxy populations explored in
\S4.  The upper envelope of points traces the minimum \sfrd\ as a
function of redshift and appears to rise by at least 5$\times$ from
$z$ = 0 to 1, a now-familiar pattern (Lilly et al.\ 1996), though this
work has led us to question the reliability of many SFR and \sfrd\
estimates.

Radio-based SFR estimates become increasingly prone to contamination
by radio-loud AGN at $z\gg\rm 1$. Unfortunately, a consequence of
removing this via a limit on $L_{\rm 1.4GHz}$ is the rejection of
luminous star-forming galaxies obeying the far-IR/radio correlation;
adopting a median $S_{\rm 1.4GHz}$ is unlikely to be better. In
addition, some redshift-limited galaxy populations defined by color
appear less well defined than first claimed (Conselice et al.\ 2006),
limiting our ability to judge the volume probed. These effects lead to
large uncertainties so while it is clear that stacking radio data is
useful, accurate SFRs for distant galaxies require precise redshifts
together with deep, multi-frequency, high-resolution radio data
($\ll$1$''$, $\sigma_{\rm 0.6GHz}\sim \sigma_{\rm 1.4GHz}\la
1\,\mu$Jy). These will facilitate identification and removal of
accretion-related emission via radio luminosity, spectral index,
brightness temperature and morphology.



\clearpage

\clearpage

\clearpage

\begin{deluxetable}{llcrr}
\tabletypesize{\scriptsize}
\tablecaption{SINGLE-COLUMN:
1.4-GHz flux densities ($S_{\rm 1.4GHz}$, in $\mu$Jy) and SFRs
determined by pixel stacking at the positions of galaxies in the samples discussed
in \S4\label{table}}
\tablewidth{0pt}
\tablehead{
\colhead{Sample\tablenotemark{a} (\S4)} & \colhead{Selection criteria} &
\colhead{N\tablenotemark{b}} & \colhead{$S_{\rm 1.4GHz}$} &
\colhead{SFR\tablenotemark{c}}
}
\startdata
UV-selected          &$z$=0.0--0.2          &136-0-0  &4.1$\pm$1.1&0.12\\
galaxies,            &$z$=0.2--0.4          &678-9-0  &11.2$\pm$0.5&3.6\\
$R_{\rm AB}$$<$24.1  &$z$=0.4--0.6          &612-12-1 &8.4$\pm$0.5&8.9\\
                     &$z$=0.6--0.8          &1,055-20-2&8.8$\pm$0.4&21\\
                     &$z$=0.8--1.0          &667-7-0  &8.5$\pm$0.5&36\\
                     &$z$=1.0--1.2          &441-7-5  &6.8$\pm$0.6&47\\
                     &$z$=1.2--1.4          &276-3-2  &5.7$\pm$0.8&59\\
                     &$z$=1.4--1.6          &30-1-1   &4.5$\pm$2.4&65\\
\tableline
EROs,                &$R$--$K_{\rm s}$=5.3--5.6  &216-14-7 &13.6$\pm$0.8&126\\
$K_{\rm s}$$<$20.5   &$R$--$K_{\rm s}$$>$5.6     &170-15-8 &11.9$\pm$0.9&111\\
$R$--$K_{\rm s}$$>$5.3&$K_{\rm s}$=17--18        &35-7-4   &20.6$\pm$2.2&194\\
                     &$K_{\rm s}$=18--19         &187-19-10&16.9$\pm$0.9&159\\
                     &$K_{\rm s}$=19--20         &140-3-1  & 6.3$\pm$1.0&59\\
                     &$K_{\rm s}$=20--21         &19-0-0   &12.9$\pm$2.8&121\\
\tableline
EROs,                &$R_{\rm AB}$--3.6=4.0--4.5&1,027-53-17& 11.8$\pm$0.4&81\\
$R_{\rm AB}$--3.6$>$4&$R_{\rm AB}$--3.6=4.5--5.0&720-29-4  &  10.6$\pm$0.4&73\\
                     &$R_{\rm AB}$--3.6=5.0--5.5&437-29-8  &  14.4$\pm$0.6&100\\
                     &$R_{\rm AB}$--3.6=5.5--6.0&137-11-2  &  16.7$\pm$1.0&115\\
                     &$R_{\rm AB}$--3.6$>$6.0   &42-8-1    &  27.7$\pm$1.8&192\\
                     &3.6$\mu$m=19--20          &411-62-15 &  25.1$\pm$0.6&174\\
                     &3.6$\mu$m=20--21          &1032-48-9 &  13.1$\pm$0.4&90\\
                     &3.6$\mu$m=21--22          &742-10-1  &  6.4$\pm$0.4&44\\
                     &3.6$\mu$m$>$22            &154-3-2   &  4.6$\pm$0.9&32\\
\tableline
24$\mu$m             &$S_{\rm 24}$=0.15--0.3\,mJy&634-65-291&23.5&1,400\\
galaxies,            &$S_{\rm 24}$=0.3--0.6\,mJy &196-61-89 &41.6&2,480\\
$S_{4.5}$$>$$S_{3.6}$&$S_{\rm 24}$=0.6--1.2\,mJy &67-34-21  &67.2&4,010\\
                     &$S_{\rm 24}$$>$1.2\,mJy    &29-19-2   &103&6,120\\
                     &$R_{\rm AB}$$<$23          &173-46-86 &38.0&2,260\\
                     &$R_{\rm AB}$=23--24        &154-39-67 &32.7&1,950\\
                     &$R_{\rm AB}$=24--25        &222-43-87 &27.5&1,640\\
                     &$R_{\rm AB}$=25--26        &151-26-69 &29.0&1,730\\
                     &$R_{\rm AB}$$>$26          &92-7-41   &21.7&1,290\\
\tableline
DRGs,                &$K_{\rm s}$$<$20.5         &80-2-1  &10.1$\pm$1.3&150\\
$J$--$K_{\rm s}$$>$2.3&                          &        &           \\
\tableline
LBGs                 &                           &107-0-28&2.0$\pm$2.3&$<$500\\
\tableline
ILLBGs               &$S_{\rm 24}$$>$0.06\,mJy   &6-1-4   &44.2$\pm$11.5&2,120\\
\enddata
\tablenotetext{a}{Units are magnitudes unless otherwise stated}
\tablenotetext{b}{Number of sources, number detected individually
(S/N $\ge$ 5) and number classified as radio-loud AGN}
\tablenotetext{c}{Radio-based SFR per object; for the $R_{\rm AB}$--3.6 EROs,
we assume $z=\rm 1.1$; for the 24-$\mu$m-selected galaxies, we assume $z=\rm 2.75$}
\end{deluxetable}

\end{document}